# A common supersolid skin covering both water and ice

Xi Zhang[1,2,a], Yongli Huang[3,a], Zengsheng Ma[3], Yichun Zhou[3,*], Weitao Zheng[4], Ji Zhou[5], and Chang Q Sun[1,*]

[1] NOVITAS, School of Electrical and Electronic Engineering, Nanyang Technological University, Singapore 639798
[2] Center for Coordination Bond and Electronic Engineering, College of Materials Science and Engineering, China Jiliang University, Hangzhou 310018, China
[3] Key Laboratory of Low-dimensional Materials and Application Technology (Ministry of Education) and Faculty of Materials, Optoelectronics and Physics, Xiangtan University, Xiangtan, 411105, China
[4] School of Materials Science, Jilin University, Changchun 130012, China
[5] State Key Laboratory of New Ceramics and Fine Processing, Department of Materials Science and Engineering, Tsinghua University, Beijing 100084, China

Abstract

Consistency in experimental observations, numerical calculations, and theoretical predictions revealed that skins of water and ice share the same attribute of supersolidity characterized by the identical H-O vibration frequency of 3450 cm$^{-1}$. Molecular undercoordination and inter-electron-pair repulsion shortens the H-O bond and lengthen the O:H nonbond, leading to a dual process of nonbonding electron polarization. This relaxation-polarization process enhances the dipole moment, elasticity, viscosity, thermal stability of these skins with 25% density loss, which is responsible for the hydrophobicity and toughness of water skin and for the slippery of ice.

TOC entry

**A supersolid skin that is elastic, hydrophobic, thermally stable, and density less slipperizes ice and toughens water surface.**

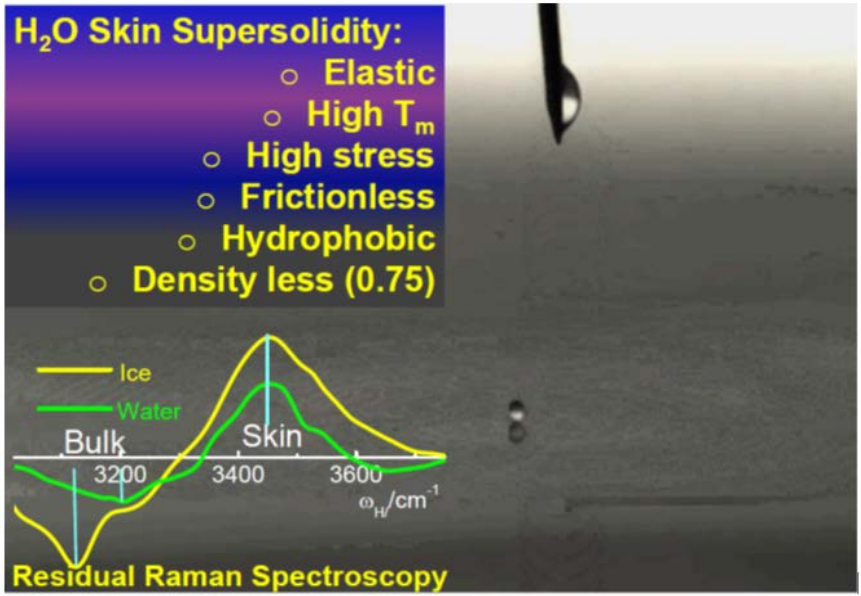

Contents

1   Introduction

Molecular undercoordination from the ideal bulk value of four makes the skin of ice slipperiest and the skin of water hydrophobic and toughest of ever known. The high lubricity of ice enabled moving heavy stones from far distance away to the Forbidden City in ancient China [1]. The high surface tension, 72 dynes/cm at 25 °C [2], toughens water skin with fascinating phenomena [3]. For instance, small insects such as a strider can walk and glide freely on water because: (i) it weighs insufficiently to penetrate the skin and (ii) the interface between its paddle and the skin of water is hydrophobic. If carefully placed on the surface, a small needle floats on water even though its density is times higher than that of water. If the surface is agitated to break up the tension, then the needle will sink quickly.

Amazingly, the skin of water ice is elastic [4], hydrophobic [5, 6], polarized [7, 8], dielectrically instable [9], thermally stable [10] with densely entrapped bonding electrons [11-14] and ultra-low mass density [15].A video clip [16] shows that a water droplet bounces continuously and repeatedly on water, which evidences the elasticity and hydrophobicity of both skins of the bulk water and the droplet.
However, correlation between the molecular coordination environment and the associated anomalies demonstrated by water and ice remains challenging despite intensive investigation since 1859 when Faraday [17] firstly proposed that a quasi-liquid skin presents lubricating ice even at temperature far below freezing point [17-19]. The slippery of ice is also perceived as pressure-promoted melting [20] and friction-induced heating [21] while the extraordinary hydrophobicity and toughness of water skin are attributed to the presence of a layer of molecules in solid state [22, 23].

In fact, ice remains slippery while one is standing still on it without involvement of pressure melting or friction heating [4, 24, 25]. According to the phase diagram of ice, pressure induces no solid-liquid transition at temperatures below -22 °C [20, 26] but slippery remains at temperatures far below this point so the mechanism of pressure melting is unlikely. On the other hand, if a liquid lubricant exists on ice, the vibration amplitudes of the skin molecules should be greater than that in the bulk ice but interfacial force microscopy measurements resolved not the expected situation [24]. The skin layer is, however, rather viscoelastic in the range from -10 to -30 °C, which evidences the absence of the liquid skin at such range of temperatures.

Furthermore, water skin is ice-like at the ambient temperatures. Sum frequency generation spectroscopy measurements and molecular dynamics (MD) calculations suggested, however, that the outermost two layers of water molecules are ordered "ice like" at room temperature [27]. At the ambient, ultrathin films of water performs like ice with hydrophobic nature [5, 23]. These facts exclude the possibilities of skin pre-melting, pressure melting, liquid skin presenting, or even the friction heating for the slippery of ice. A small-angle scattering of X-ray measurements of water structure at temperatures of 7, 25, and 66 °C under atmospheric pressure have resolved the core-shell structure of water with a skin of ~0.12 nm thick, agreeing with the TIP4P/2005 force-field calculation derivative of a 0.04–0.12 nm thick shells [28].

Does a liquid skin form on ice or a solid skin covers the liquid water? One has to address this paradox from the perspective of skin O:H-O bond relaxation. This work shows a solution to this mystery based on our recent progress [29]. We show that molecular undercoordination shortens the H-O bond and lengthens the O:H nonbond by repulsion between electron pairs on adjacent oxygen ions [29], which polarize the O:H nonbond electrons to form a common supersolid skin characterized by the $\omega_H = 3450$ cm$^{-1}$ on both water and ice. The skin supersolidity is responsible not only for the high elasticity, hydropphobisity, thermal stability, and toughness of water skin but also for the slippery of ice.

2    Principle and predictions
2.1    Principle: O:H-O bond cooperativity

Figure 1 illustrates the O:H-O bond configuration with short-range interactions [29]. The O:H-O bond is segmented into the longer-and-softer O:H nonbond (left-hand side with ":" representing the lone pair of electrons) and the shorter-and-stiffer H-O polar-covalent bond (right-hand side). The O:H interacts with a van der Walls like (vdW-like) potential that contains the electrostatic interaction between the H$^+$ and the O$^{2-}$ ions with binding energy at 0.1 eV level. The H-O bond exchange interaction has a ~4.0 eV cohesive energy. The O:H-O serves as a pair of asymmetric, H-bridged oscillators coupled strongly with the Coulomb repulsion between electron pairs (green pairing dots) on adjacent O atoms. The diameters of the springs represent the strengths of the respective interactions.

Recent advancement revealed that the Coulomb coupling and the external excitation (molecular undercoordination [29], mechanical compression [30], and thermal excitation [31]) drive the O:H-O bond to relax cooperatively. O atoms always dislocate in the same direction but by different amounts along the O:H-O bond with H being the coordination origin [30].The softer O:H relaxes more than the stiffer H-O does under excitation. The undercoordinated molecule, the H-O contraction and the O:H elongation dominate the relaxation [29].

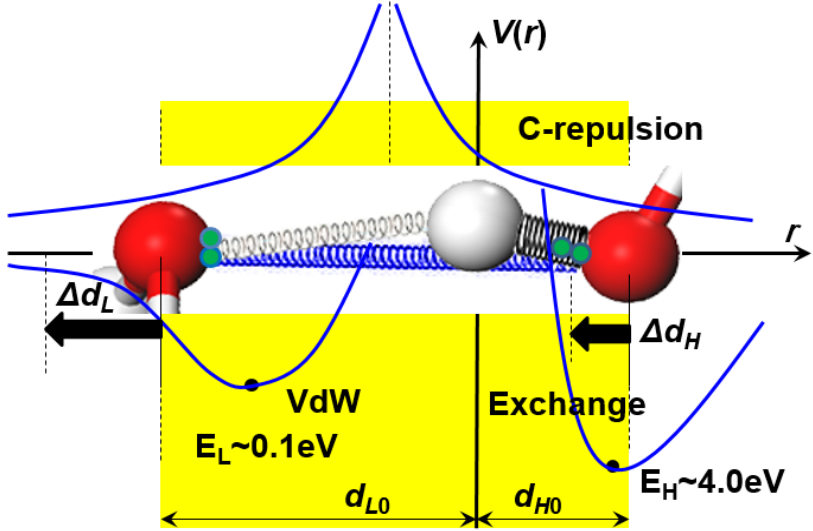

Figure 1 Asymmetric, local, and short-range potentials for the segmented O:H-O bond. Interactions include the O:H nonbond (vdW-like, left-hand side) force, the H-O exchange (right-hand side) interaction, and the O---O Coulomb repulsion (C-repulsion) [32]. Molecular coordination number (CN) reduction and the C-repulsion shorten the $d_H$ spontaneously and lengthen the $d_L$, elongating the O---O and reducing the mass density [29].

2.2 Predictions

2.2.1 Skin phonon cooperative relaxation and $T_c$ elevation

The concept of supersolidity is adopted from the superfluidity of solid $^4$He at mK temperatures. The skins of $^4$He fragments are highly elastic and frictionless with repulsive force between them at motion [33]. Molecular undercoordination and inter-oxygen repulsion relax the O:H-O bond cooperatively and ubiquitously to skins, defects, hydration shells, and ultrathin films of water ice. Bond relaxation and the associate energetics, localization, entrapment, and polarization of electrons mediate the performance of the skins, clusters, or thin films of water and ice, in the following ways [29]:

$$\left.\begin{array}{c} T_C \\ \Delta E_{1s} \\ \Delta \omega_x \end{array}\right\} \propto \left\{\begin{array}{c} E_H \\ E_H \\ \sqrt{E_x/\mu_x}/d_x \propto \sqrt{Y_x d_x/\mu_x} \end{array}\right.$$

(1)

where $E_x$ and $d_x$ are the cohesive energy and bond length, respectively, of the respective bonding segment (x = H or L). The $\mu_x$ is the reduced mass of the respective H$_2$O:H$_2$O and the H-O oscilator. The critical temperature ($T_C$ except for evapoaration that is dominated by the O:H dissociation energy $E_L$) for phase transition is proportional to the $E_H$. Theoretical reproduction of pressure dependence of the $T_C$ for ice VII-VIII phase transition verified that the $E_H$ determines the $T_C$ [30] with an estimation of $E_H$ = 3.97 eV for bulk ice. is Ehe O 1s binding energy shift $\Delta E_{1s}$ is proportional to the $E_H$. The phonon frequency shift ($\Delta \omega_x$) is proportional to the square-root of the respective bond sfiffness ($Y_x d_x$) that is the product of the bond length and elastic modulus. The $Y_x$ is proportional to the binding energy density, $Y_x \propto E_x d_x^{-3}$ [30].

### 2.2.2  Skin electronic entrapment and polarization

According to the tight-binding theory [34], H-O bond stiffening deepens the interamolecuar potential well, which shifts the $E_{1s}$ from that of an isolated O atom $E_{1s}(0)$ to the $E_{1s}(z)$ in the form of:

$$\frac{\Delta E_{1s}(z)}{\Delta E_{1s}(4)} = \frac{E_{1s}(z) - E_{1s}(0)}{E_{1s}(4) - E_{1s}(0)} = \frac{E_H(z)}{E_H(4)}$$

(2)

The $\Delta E_{1s}$ is proportional to the $E_H$, as the $E_L \sim 0.1$ eV is negligibly small.

Two processes of polarization of the nonbonding electrons take place at the skin [29]. One is the lone pair polarization by the densely entrapped H-O bonding electrons of the same oxygen atom and the other is the O---O repulsion caused by electron gain of undercoordinated molecules. Repulsion lengthens the O---O and polarizes the lone pairs to raise the local dipole moment, which decreases the vertical bound energy (work function) of electrons at skin.

### 3  Experimental and numerical approaches

#### 3.1  Residual O:H-O length spectroscopy (RLS)

Firstly, we calculated the O:H and the H-O length, and the respective vibration frequencies using the DFT-MD and the dispersion-corrected DFT packages. In computation, we adopted the conventionally used super-cell [35] as shown in Figure 2, with and without a vacuum slab being inserted to approximate the absence and presence of the liquid-vapor interface or the skin that contains free H-O radicals. Calculations were focused on the O:H-O cooperative relaxation in length, phonon stiffness, and the skin charge accumulation.

MD calculations were performed using the Forcite code with COMPASS force field [36]. Ice interface is relaxed in NPT ensemble at 180 K for 100 ps to obtain equilibrium. The time step is 0.5 fs. Nose-hoover thermostat with Q ratio of 0.01 is adopted to control the temperature. Dispersion-corrected DFT structural optimizations of ice surface were performed using Dmol$^3$ code based on the PBE functional [37] in the general gradient approximation and the dispersion-corrected Tkatchenko-Scheffler scheme [38] with the inclusion of hydrogen bonding and vdW interactions. We used all-electron method to describe the wave functions with a double numeric and polarization basis sets. The self-consistency threshold of total energy was set at $10^{-6}$ Hartree. In the structural optimization, the tolerance limits for the energy, force, and displacement were set at $10^{-5}$ Hartree, 0.002 Hartree/Å and 0.005 Å, respectively.

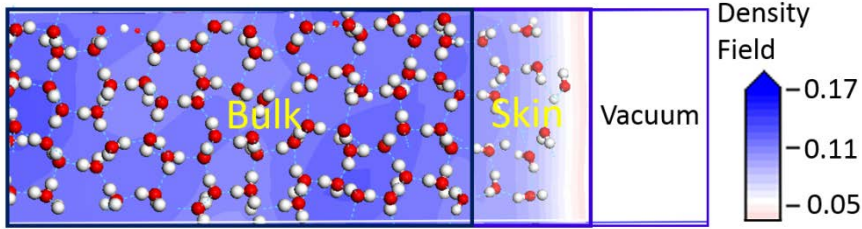

Figure 2 Schematic illustration of the super-cell of water with an insertion of a vacuum slab represents the supersolid skin of ice (200 K). There are regions, from left to right, of the bulk, the skin, and the vacuum. The skin contains undercoordinated molecules and the free H-O radicals. Color scales the MD-derived mass density field in the unit cell. $H_2O$ molecule shrinks its size but inflates their separations in the skin and hence lowers the local mass density.

3.1  Residual phonon spectroscopy (RPS)

In order to verify our predictions, we analyzed the Raman spectra collected by Donaldson and co-workers [4]. They probed the phonon frequency using Raman spectroscope from water and ice, respectively, at emission angles of 87° and 0° with respect to the surface normal using a 355 nm laser beam. We obtained the residual phonon spectrum (RPS) by differencing these two spectra upon the spectral peak areas being normalized and background corrected. Hence, we can discriminate the characteristic peaks of the skin H-O bond of molecules with lowest CN from that of the bulk with highest CN. The commonly shared area in the raw spectra being not of immediate interest has been removed from the processing.

3.3  Skin stress and viscosity

We also calculated the surface tension and thickness dependent viscosity using the classical method briefed as follows. The difference between the stress components in the directions parallel and perpendicular to the interface defines the surface tension γ [39, 40],

$$\gamma = \frac{1}{2}\left(\frac{\sigma_{xx}+\sigma_{yy}}{2} - \sigma_{zz}\right) \cdot L_z$$

(3)

where $\sigma_{xx}$, $\sigma_{yy}$, and $\sigma_{zz}$ are the stress tensor element and $L_z$ the length of the super-cell. The surface shear viscosity $\eta_s$ is correlated to the bulk stress $\sigma$ as [41, 42]:

$$\eta_s = \frac{V}{kT} \int_0^\infty \langle \sigma_{\alpha\beta}(0)\sigma_{\alpha\beta}(t) \rangle dt \tag{4}$$

where the $\sigma_{\alpha\beta}$ denote the three equivalent off-diagonal elements of the stress tensors. The bulk viscosity $\eta_v$ depends on the decay of fluctuations in the diagonal elements of the stress tensor:

$$\eta_v = \frac{V}{kT} \int_0^\infty \langle \delta\sigma(0)\delta\sigma(t) \rangle dt$$
$$\delta\sigma = \sigma - \langle \sigma \rangle \tag{5}$$

Based on these notations, we calculated the γ using MD method to derive the stress tensors first. We also use the auto-correlation functions of stress tensors to calculate the $\eta_s$ and $\eta_v$ according to eq (4) and (5).

4 Results and discussion

4.1 Skin bond-electron-phonon attributes

4.1.1 Mass density loss due to $d_x$ relaxation

Figure 3 features the residual length spectra (RLS) for the MD-derived O-H and O:H bond of ice. We obtained the RLS by subtracting the length spectrum of the unit cell without the skin from that with a skin. Insets are the calculated raw spectra. The RLS turns out that the H-O bond contracts from the bulk value of ~1.00 to ~0.95 at the skin, which is associated with O:H elongation from ~1.68 to ~1.90 Å with high fluctuation. This cooperative relaxation results in a 6.8% $d_{OO}$ elongation or a 20% volume expansion. The peak of $d_H = 0.93$ Å and the broad $d_L$ peak correspond to the even undercoordinated H-O radicals that are associated with 3650 cm$^{-1}$ vibration frequency [29].

The skin O:H nonbond expands more than the skin O-H bond contracts, which lengthens the O---O distance and lowers the mass density [15, 43-45]. As reported in [15, 32], the skin $d_{OO}$ was measured to expand by 5.9 ~ 10.0 % at the ambient. Likewise, a 4.0% and 7.5% volume expansion has been resolved for water droplets confined in the 5.1 and 2.8 nm sized hydrophobic TiO$_2$ pores, respectively [46]. MD calculations also revealed that the H-O bond contracts from 0.9732 Å at the center to 0.9659 Å at the skin of a water droplet containing 1000 molecules [47]. These findings evidence the predicted mass density loss of the hydrophobic water skin.

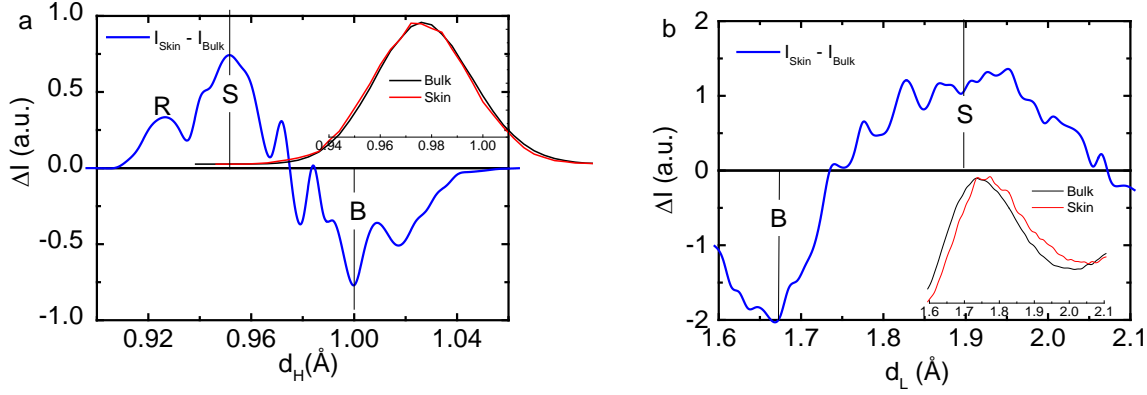

Figure 3 MD-derived RLS reveals (a) $d_H$ contraction from the bulk value (B valley) of ~1.00 to ~0.95 Å for the skin (S) and to 0.93 Å for the H-O free radicals (R) and (b) $d_L$ elongation from bulk value (B valley) of ~1.68 to ~1.90 Å (with high fluctuation). Insets show the raw data of calculations.

With the known O-H and H:O bond length relaxation and the tetrahedrally-coordinated structure [29], we obtained the size $d_H$, separation $d_{OO}$, and mass density ρ of molecules packing in water ice in the following relationships with the $d_{H0}$ and the $d_{L0}$ being the references at 4°C [32],

$$\begin{cases} d_{oo} = 2.6950\rho^{-1/3} & (Molecular\ separation) \\ \dfrac{d_L}{d_{L0}} = \dfrac{2}{1+exp[(d_H - d_{H0})/0.2428]}; & (d_{H0} = 1.0004\ and\ d_{L0} = 1.6946\ at\ 4\ °C) \end{cases}$$

(6)

With the measured $d_{OO}$ of 2.965 Å [15] as input, this relation yields the segmental lengths of $d_H$ = 0.8406 Å and $d_L$ = 2.1126 Å, which turns out a 0.75 g·cm$^{-3}$ skin mass density. In comparison, the MD derivatives in Figure 3 turn out a 0.83 g·cm$^{-3}$ skin mass density. Within the tolerance, both densities are much lower than the bulk value of 0.92 g·cm$^{-3}$ for ice. Indeed, the mass density of both skins suffers loss due to molecular undercoordination. Table 1 lists the $d_{OO}$, $d_x$, ρ, and $\omega_x$ for the skin and the bulk of water and ice in comparison to those of ice at 80 K and water dimers with the referenced data as input.

Table 1 Experimentally-derived skin supersolidity ($\omega_x$, $d_x$, ρ) of water and ice. With the known measured $d_x$ and $\omega_x$ as input, eq (1) and (6) yield the other parameters.

|  | Water (298 K) |  | Ice (253K) | Ice(80 K) | Vapor |
|---|---|---|---|---|---|
|  | bulk | skin | bulk | Bulk | dimer |
| $\omega_H$(cm$^{-1}$) | 3200[4] | 3450[4] | 3125[4] | 3090[31] | 3650[48] |

| $\omega_L$(cm$^{-1}$)[31] | 220 | ~180[29] | 210 | 235 | 0 |
| --- | --- | --- | --- | --- | --- |
| $d_{OO}$(Å) [32] | 2.700[49] | 2.965[15] | 2.771 | 2.751 | 2.980[15] |
| $d_H$(Å) [32] | 0.9981 | 0.8406 | 0.9676 | 0.9771 | 0.8030 |
| $d_L$(Å) [32] | 1.6969 | 2.1126 | 1.8034 | 1.7739 | ≥2.177 |
| $\rho$(g·cm$^{-3}$) [32] | 0.9945 | 0.7509 | 0.92[50] | 0.94[50] | ≤0.7396 |

### 4.1.2 Identical phonon frequency of water and ice

Figure 4(a, b) features the calculated RPS for ice in comparison to (c) the measured RPS of the $\omega_H$ for both water and ice with insets showing the original raw Raman spectra [4]. The valleys of the RPS represent the bulk feature while the peaks the skin and the radical attributes. Proper offset of the calculated RPS is necessary as the MD code overestimates the intra- and inter-molecular interactions [31].

As expected, the $\omega_L$ undergoes a redshift while the $\omega_H$ a blue with splitting into three components. The $\omega_H$ shifts arise from the stiffening of the skin H-O bonds (S) and the free H-O radicals (R). The $\omega_L$ redshift arises from O---O repulsion and polarization. The polarization in turn screens and splits the intramolecular potential, which results in another $\omega_H$ peak (denoted P as polarization) with frequency being lower than that of the bulk valley (B).

Most strikingly, the measured RPS shows that skins of water and ice share the same $\omega_H$ value of 3450 cm$^{-1}$, which indicates that the H-O bond in both skins are identical – the same length and the same energy as the expression of the $\omega_H \propto (E_H/d_H^2)^{1/2}$ is unique. The skin $\omega_L$ of ice may deviate from that of water because of the extent of polarization though experimental data is lacking. Nevertheless, the skin $\omega_H$ stiffening agrees with the DFT-MD derivatives that the $\omega_H$ shifts from ~3250 cm$^{-1}$ at a 7 Å depth to ~3500 cm$^{-1}$ of the 2 Å skin of liquid water [35]. Therefore, neither an ice skin forms on water nor a liquid skin covers ice, instead, an identical supersolid skin covers both.

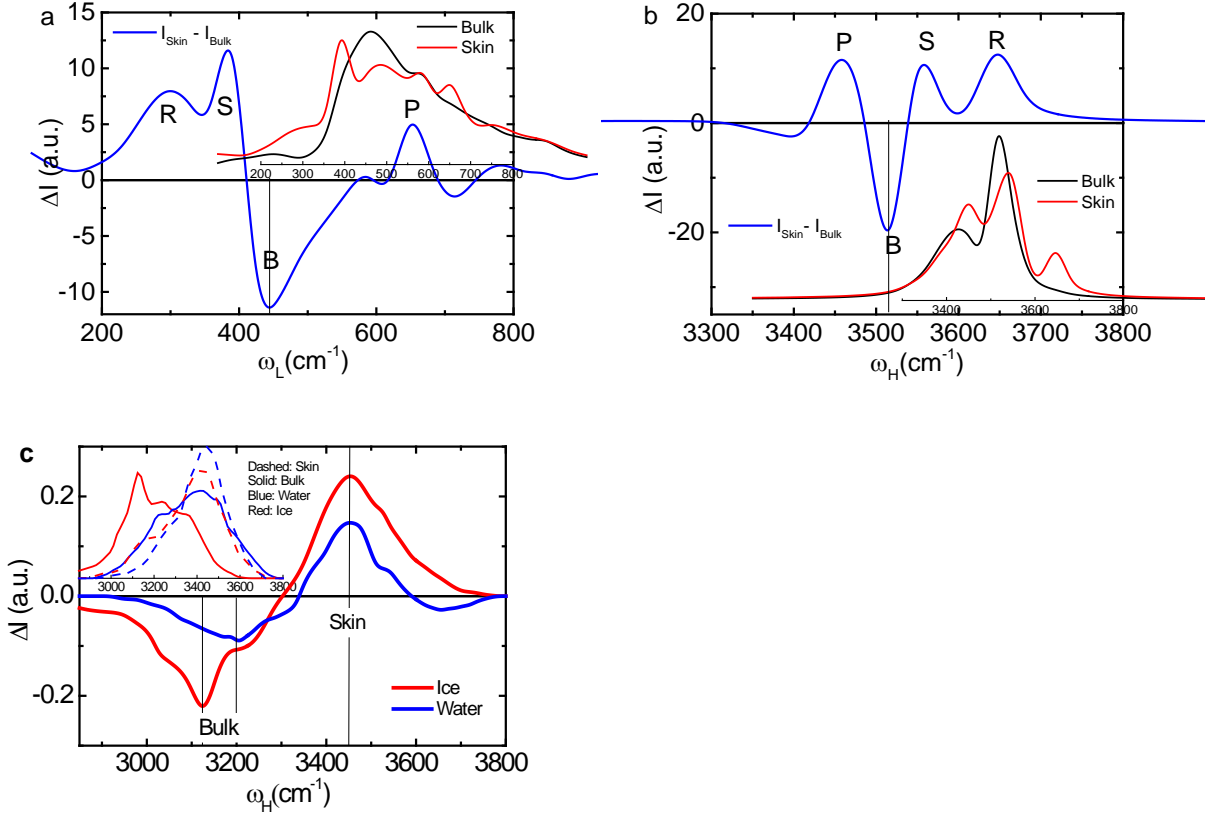

Figure 4 RPS of the MD-derived (a) $\omega_L$ and (b) $\omega_H$ of ice and (c) the measured $\omega_H$ of water (at 25 °C) and ice (at -20 and -15 °C) [4]. Insets show the original spectra. Dashed lines in inset of (c) represent the skin spectra and solid lines the bulk for ice (red) and water (blue). Calculation results show that the $\omega_L$ undergoes a redshift while the $\omega_H$ splits into three. Features S and R correspond to the $\omega_H$ of the skin (S) and of the free H-O radicals, respectively; the P component arises from the screening and splitting of the crystal potentials by the polarization of the undercoordinated O:H in numerical derivatives. Water and ice skins share the same $\omega_H$ of 3450 cm$^{-1}$, which clarifies that the length and energy of the H-O bond in the skins are identical disregarding temperature and phase difference. The intensity difference arises from scattering by ice and water.

### 4.1.3 Electronic entrapment and polarization

Table 2 features the DFT-derived Mulliken charge at the skin and the bulk of water. The net charge of a water molecule increases from the bulk value of 0.022 to -0.024 e at the skin. The densification and entrapment of bonding electrons polarize the nonbonding electrons. As it has been discovered using an ultra-fast liquid jet vacuum ultra-violet photoelectron spectroscopy [7], the bound energy for an nonbonding electron in solution changes from a bulk value of 3.3 eV to 1.6 eV at the water skin. The bound energy of nonbonding electrons, as a proxy of work function and surface polarization, decreases further with molecule cluster size.

Table 2  DFT-derived charge polarization at the skin of ice. Negative sign represents net electron gain.

|     | Skin   | Bulk   |
| --- | ------ | ------ |
| $q_O$ | -0.652 | -0.616 |
| $q_H$ | 0.314  | 0.319  |
| Net | -0.024 | 0.022  |

## 4.2 Skin thermodynamics

### 4.2.1 High thermal stability

The supersolid skin is thermally stable because $T_m \propto E_H$. According to Eq.(1), the shortening of the $d_H$ raises the $E_H$. Therefore, it is not surprising that water skin performs like ice at room temperature and that the monolayer water melts at temperature about 315 K [10]. It is also clear why water droplets encapsulated in hydrophobic nanopores [51] and point defects [52, 53] are thermally more stable than the bulk water. Topological defects mediate the melting of bulk ice by preserving the coordination environment of the tetrahedral-coordinated network. Such defects form a region with a longer lifetime than the ideal bulk [53]. Another evidence for the skin thermal stability is that the critical temperature for transiting the initial contact angle of a droplet to zero is curvature dependent [54]. Transition happens at 185, 234, and 271 °C for water droplets on quartz, sapphire, and graphite with initial contact angles of 27.9, 64.2, and 84.7°, respectively.

### 4.2.2 Elasticity, viscosity, and hydrophobicity elevation

Table 3 features the MD-derived thickness-dependent $\gamma$, $\eta_s$, and $\eta_v$ of ice films. Reduction of the number of molecular layers increases all values of the $\gamma$, $\eta_s$, and $\eta_v$. The O:H-O cooperative relaxation and the associated entrapment and polarization enhance the surface tension to reach the values of 73.6 dyn/cm for 5 layers, approaching the measured 72 dyn/cm at 25 °C.

Table 3 Thickness-dependent surface tension and viscosity of ice skin.

| Number-of-layer | 15 | 8 | 5 |
| --- | --- | --- | --- |
| $\gamma$ (dyn/cm) | 31.5 | 55.2 | 73.6 |
| $\eta_s$ (dyn·s/cm$^2$) | 0.007 | 0.012 | 0.019 |
| $\eta_v$ (dyn·s/cm$^2$) | 0.027 | 0.029 | 0.032 |

The negative charge gain and the nonbonding electron polarization on oxygen provide electrostatic repulsive force lubricating ice (see Table 2). Measurements verified indeed the presence of the repulsive forces between a hydrated mica-tungsten contacting pair at 24 °C [55]. Such repulsive interactions appear at 20-45% relative humidity (RH). The repulsion corresponds to an elastic modulus of 6.7 GPa. Monolayer ice also forms on graphite surface at 25% RH at 25 °C [56]. These experimental findings and the present numerical derivatives evidence the presence of the supersolidity with repulsive forces because of bonding charge densification, surface polarization, and the elevated $T_m$.

4.3  Skin supersolidity slipperizes ice

Because of the dual-process of polarization, the skin forms such an amazing supersolid phase that slipperizes ice. Skin supersolidity may mimic the superfluidity of solid $^4$He [33] and water droplet flowing in the carbon nanotubes [57]. The H-O contraction, core electron entrapment, and polarization yield the high-elasticity, self-lubrication, and low-friction of ice and the hydrophobicity of water surface as well, of which the mechanism is the same to that of metal nitride [58, 59] surfaces with electron lone pair coming into play. Nanoindentation measurements revealed that TiCrN and GaAlN surfaces could reach 100% elastic recovery under a critical indentation load of friction (< 1.0 mN) at which the lone pair breaks. The friction coefficient is in the same order to ice (0.1)[25]. The involvement of lone pairs and the skin bond contraction make the nitride skins more elastic and slippery under the critical load, which also explains why the flow rate of water droplet in carbon nanotubes increases with reduction of the tube diameter [57].

5  Conclusion

Consistency in experimental measurements, numerical calculations, and our predictions based on the "molecular undercoordination induced O:H-O cooperative relaxation and polarization" confirms the following:

1) Molecular undercoordination shortens the $d_H$ from ~1.0 to 0.84 Å and lengthens the $d_L$ from ~1.75 to 2.11 Å through inter-oxygen repulsion, which lowers the mass density to 0.75 g·cm$^{-3}$ of the skin from the values of 0.92/1.0 for bulk ice/water.
2) The shortening of the H-O bond stiffens the H-O phonon from 3150~3200 cm$^{-1}$ to an identical value of 3450 cm$^{-1}$ for the skin of both water and ice, raises the melting temperature from 273 to ~315 K, and entraps the O 1s binding energy from the bulk value of 536.6 to 538.1 eV.
3) A dual-process of nonbonding electron polarization enhances the elasticity, viscosity, and dipole moment of these skins.
4) Neither liquid skin forms on ice nor ice skin covers water, instead, a supersolid skin presents not only lubricating ice toughening water skin.

Acknowledgement

Financial support received from NSF (Nos.: 21273191, 1033003, and 90922025) China and RG29/12, MOE,

FOOT NOTES

[a] These authors contribute equally.

*To whom correspondence should be addressed: ecqsun@ntu.edu.sg; CQ is associated with honorary appointments at rest affiliations.